\documentclass[final]{IEEEtran}

\usepackage{amsmath,amssymb,amsfonts}
\usepackage{algorithmic}
\usepackage{xcolor}
\usepackage{cite}

\usepackage{times}
\usepackage{textcomp}
\usepackage{graphicx}
\usepackage{amssymb}
\usepackage{url,hyperref}
\usepackage{enumitem}
\usepackage{type1cm}
\usepackage{lettrine}
\usepackage{booktabs}
\usepackage{multirow}
\usepackage{supertabular}
\usepackage{tabularx}

\begin{document}

\title{Phish-Defence: Phishing Detection Using Deep Recurrent Neural Networks}

\author{
\IEEEauthorblockN{Aman Rangapur, 
			Tarun Kanakam and Dhanvanthini P
		}	

\thanks {Aman Rangapur is with the School of Computer Science and Engineering, VIT-AP University, Andhra Pradesh, India, 522237. Email: amanrangapur@gmail.com} %\protect\\

\thanks {Tarun Kanakam is with the School of Computer Science and Engineering, VIT-AP University, Andhra Pradesh, India, 522237, Email: kanakamtarunkumartk@gmail.com} %\protect\\ 

\thanks{Ajith Jubilson is with the School of Computer Science and Engineering, VIT-AP University, Andhra Pradesh, India, 522237, Email: ajith.jubilson@vitap.ac.in} %\protect\\
}

\maketitle
\begin{abstract}
In the growing world of the internet, the number of ways to obtain crucial data such as passwords and login credentials, as well as sensitive personal information has expanded. Page impersonation, often known as phishing, is one method of obtaining such valuable information. Phishing is one of the most straightforward forms of cyber-attack for hackers, as well as one of the simplest for victims to fall for. It can also provide hackers with everything they need to access their targets’ personal and corporate accounts. Such websites do not provide a service but instead gather personal information from users. In this paper, we achieved state-of-the-art accuracy in detecting malicious URLs using recurrent neural networks. Unlike previous studies, which looked at online content, URLs, and traffic numbers, we merely look at the text in the URL, which makes it quicker and catches zero-day assaults. The network has been optimized so that it may be utilized on tiny devices like Mobiles, Raspberry Pi without sacrificing the inference time.
\end{abstract}

\section{Introduction}
\lettrine[lines=2]{T}{}he rapid growth and adoption of new technologies, such as smart devices and 5G connectivity, has resulted in the rapid growth of internet-based services. Some of the services are essential for day-to-day activities. The increasing number of people who utilize web-based services demonstrates their popularity over the years. The goal is to increase the availability and accessibility of web-based services used frequently so that users do not face any inconvenience\cite{ar7,ar8}. Nonetheless, the unlimited access and usage of web-based services or technology on the internet offers potential for cyber-attacks because of non-uniform cyberspace regulations or processes \cite{ar6}.

There has been incredible growth in the number of illegal activities with increased malicious and destructive content in the networks over the past several decades. According to statistics, financial transactions, online gaming services, and social media are among the most popular web-based services with a large user base. These services are highly prone to cyber assaults. The way to protect the user from cyberattacks is called Cybersecurity \cite{ar1}. Cybersecurity is responsible for safeguarding, prohibiting, and recovering all the assets that use the internet from cyber assaults. 

Phishing is a social-engineering-based cyber-attack that aims to steal personal data such as credit card information, login credentials, and passwords. Most of the time, the hacker registers a counterfeit domain address, meticulously creates a website that resembles the organization's actual website, then sends a mass email to thousands, instructing recipients to click on a link on the fake website. These assaults primarily result in the loss of crucial data, financial loss, intellectual property loss, and deterioration of trust and national security \cite{ar3,ar4}. Several cybersecurity professionals and researchers have proposed and built a variety of anti-phishing techniques for identifying phishing websites \cite{ar9}. Elmer Lastdanger proposed in his study that ‘‘Phishing is a scalable act of deception whereby impersonation is used to obtain information from a target’’ \cite{ar2}.

Machine learning is a field within computer science that differs from traditional computing. In traditional computing, algorithms follow the rules and get executed, whereas Machine Learning algorithms allow computers to train on data and use statistical analysis to output the value. In addition, neural networks and genetic algorithms make the code very efficient. Neural networks are a means of machine learning, in which a computer learns to perform some task by analyzing training data; it's a process that mimics how the human brain operates. A neural network generates the best possible result without redesigning the output criteria. 

The motivation of this research is to obtain state-of-the-art accuracy using Recurrent Neural Networks, which is efficient in detecting malicious websites. Furthermore, our model has been optimized and can be utilized on tiny devices like Raspberry Pi without sacrificing the inference time.

\section{Background study}
The constant development of web services and e-commerce platforms has encouraged many phishers and criminals to develop new ways to exploit and deceive novice users into sending their financial information \cite{article10,article11}.

\begin{figure}
\centering
\includegraphics[width=0.5\textwidth]{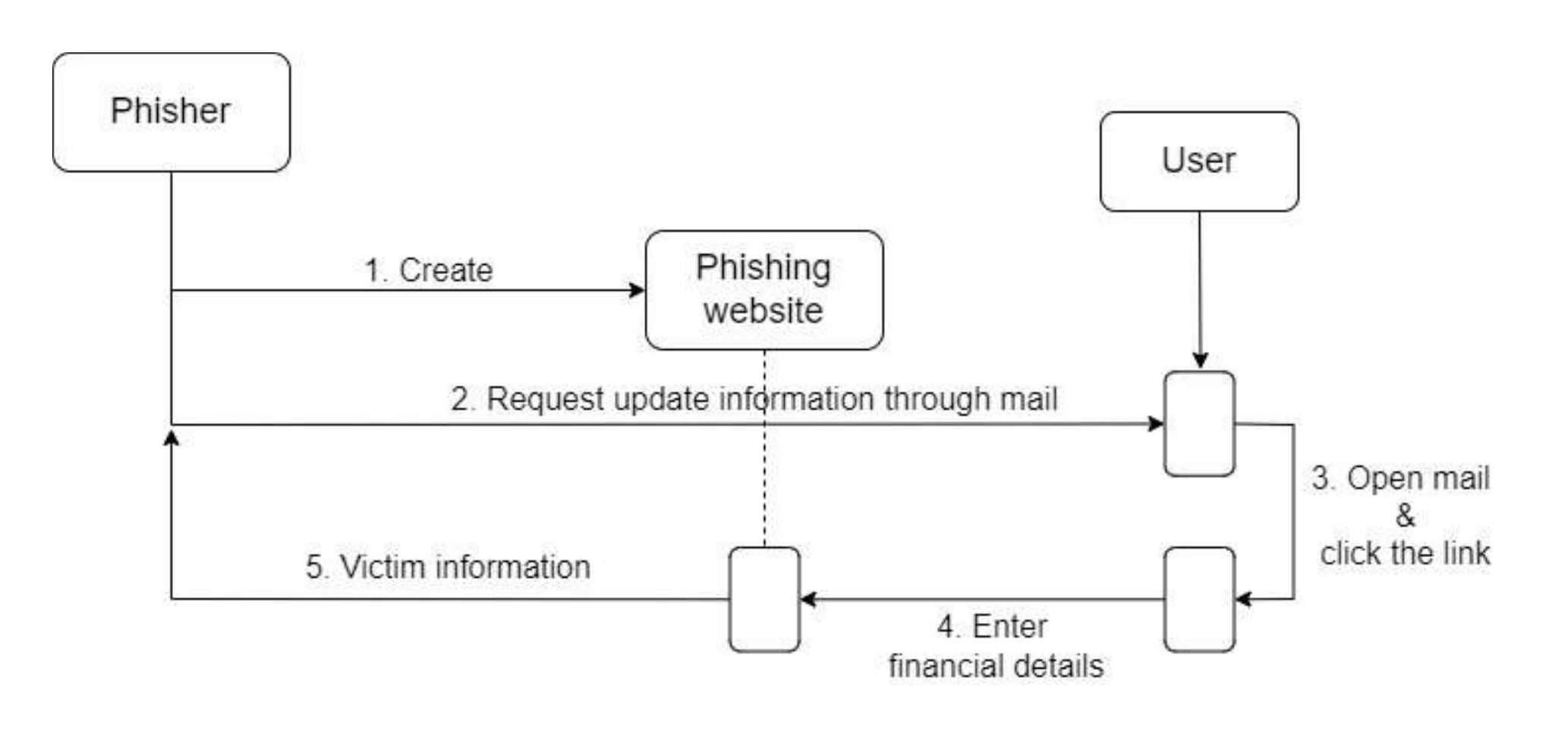}
\caption{Life cycle of a web phishing attack.}
\label{FIG:Life_cycle_of_a_web_phishing_attack}
\end{figure}

Phishing is by far the most common cybercrime as of 2022, with the FBI's Internet Crime Complaint Centre reporting more than twice as many phishing cases than any other kind of cybercrime. In a web phishing attack using email, the phisher deceives web users by developing a fake website to steal the financial and personal information of the users. Web phishing attacks can be achieved in many ways \cite{article11,article14}. Initially, the phisher makes a phishing website, which looks like the original website in its appearance. Eventually, many emails containing hyperlinks to the phishing website are continually sent to users by the phisher, which requests validating or updating credentials and financial information to deceive the victims \cite{a32}. Finally, the target victims are redirected to the phishing website when the hyperlinks are clicked. When the victim enters the information through the phishing website, the phisher can fully control the victim's financial information. As a result, many financial and identity thefts can be executed after successful web phishing attacks \cite{article11,article12,article14}. Online social networks, emails, messengers, blogs, forums,  mobile applications, and voice over IP (VoIP) are different communication platforms. Phishing attacks may target these communication systems. Figure \ref{FIG:Life_cycle_of_a_web_phishing_attack} shows the  life cycle of a web phishing attack.

Phishing attempts against software-as-a-service (SaaS), webmail users were the most common in 2021. During the COVID-19 pandemic, the number of phishing attacks increased significantly. Approximately 70\% of healthcare facilities with less than 500 employees are targeted, indicating that these smaller facilities have probably weaker security systems due to low-security budgets. It's also been observed that 75\% of phishing websites use SSL (Secure Sockets Layer) security, which means that a website that uses the SSL protocol is not always legitimate. The countermeasures in tackling phishing attacks are categorized as follows.

\subsection{Traditional methods}
\subsubsection{Blacklisting and whitelisting}
Blacklisting and whitelisting are phishing URL lists in which the specified URL is compared to a preset phishing URL. The disadvantage of this methodology is that the blacklist cannot cover all phishing sites because a newly created phishing site takes a long time to get updated in the list. This delay in adding the suspect site to the list might be enough for the attackers to achieve their goals \cite{ar7,article10}. Whitelisting methods are contrary to blacklisting methods. It determines if a URL is legitimate by comparing it to the “whitelist” Database. The “whitelist” Database mainly contains a list of well-known legitimate URLs \cite{article11}.

\subsubsection{Visual Similarity}
By analyzing the appearance of sites, Visual Similarity techniques may distinguish phishing sites from genuine ones. Attackers embed images, Flash, ActiveX, and Java Applet instead of HTML text to avoid phishing detection. As enrollment items appear on phishing site webpages, visual similarity-based detection algorithms may immediately distinguish them. It classifies URLs based on visual similarity strategies using HTML Document Object Model tree, CSS similarity, visual perception, visual characteristics, pixel-based, and hybrid approaches \cite{ar10}.

\subsection{Non-Traditional methods}
\subsubsection{Machine Learning and Deep Learning}
To create phishing detection models, traditional machine learning methods (also known as shallow learning) such as Support Vector Machines (SVM), Decision Trees (DT), and Random Forests (RF) can be used.

The Hybrid Ensemble Feature Selection (HEFS) presented by Chiew K. L. et al. \cite{ar12} is a feature selection structure for a machine learning-based phishing detection system. In the first step of HEFS, an original Cumulative Distribution Function gradient (CDF-g) calculation is used to generate critical feature sub-sets, which are then used as an input to a data perturbation collection to generate optional feature sub-sets. The subsequent stage infers pattern features from the optional feature subsets by utilizing a function disturbance set. Compared to SVM, Naive Bayes, C4.5, JRip, and PART classifiers, HEFS performs better when coordinated with the Random Forest classifier.

Deep Learning, which comes under machine learning, focuses on building deep networks with various layer types such as convolutional, pooling, dropout, and fully connected layers. Deep learning-based phishing detection models are built using algorithms like Convolutional Neural Networks (CNN), Recurrent Neural Networks (RNN), Autoencoders.

Wei W. et al. \cite{ar13} suggested a deep neural network with convolutional layers for identifying phishing sites based solely on the URL address content. Despite previous research that looked at URLs, traffic statistics, and online content, they only looked at the URL text. The network they demonstrated had been appropriately updated, so it can now be used even on mobile phones without affecting its performance. As a result, the approach is faster and can detect zero-day threats.

\subsubsection{Heuristic based methods}
In Heuristic-based methods, minor characteristics of a website are extracted to determine if the URL is malicious or legitimate. In contrast to blacklisting approaches, heuristic-based solutions may detect new phishing websites on a regular basis\cite{article9}. Intelligent machine learning classifiers are trained using certain phishing and authentic websites as training data and then utilized to detect newly developed phishing websites accurately \cite{article14}.

Rao R. S. et al. \cite{ar11} developed a heuristic technique with TWSVM (twin support vector machine) classifier to detect malicious enrolled phishing websites. Moreover, websites are facilitated on arrangement servers. Their research has focused on detecting phishing sites hosted on arrangement domains by inspecting the sign-in page and the website's main page. The hyperlink and URL – based characteristics are used to identify maliciously enlisted phishing websites.

\subsubsection{Hybrid Techniques}
A combination of different approaches is used to create a better model in terms of accuracy and precision.
A DTOF-ANN ("Decision Tree and Optimal Features based Artificial Neural Network"), which is a neural-network phishing identification model based on a decision tree and ideal feature selection, was proposed by Zhu E. et al. \cite{ar14} to tackle over-fitting. The initial step was to improve the traditional K-medoids clustering technique by selecting a consistent number of introduction centers to eliminate copy dots in popular datasets. In Addition to DTOF-ANN, optimal feature selection based on innovative features, such as decision trees and neighborhood search techniques, is meant to eliminate the ineffective and undesirable parts. Finally, the optimal architecture of the neural network classifier is constructed by modifying boundaries and training the neural network classifier with the specified ideal features.

\section{Related Work}

Researchers have studied phishing attacks repeatedly, but most of them were not entirely perfect. Despite getting solutions and accuracy, it required complicated calculations, which made it difficult to use and took a good amount of inference time to detect.

Using a variety of characteristics, da Silva et al. \cite{ar16} developed a phishing prediction model. The suggested model assesses static characteristics like Keywords and patterns in the phishing URL. A qualitative examination of these characteristics that do not define elements such as relationships and similarities between qualities was undertaken in addition to the quantitative data in the research.

A neural network model based on decision trees and optimal feature selection was suggested by Zhu et al. \cite{ar14}. Using the classic K-medoids clustering process, they created an incremental selection strategy to eliminate duplicate points from public datasets. An optimum feature selection method was created to exclude the negative and useless traits.
Tan et al. \cite{ar17} developed graph theory-based anti-phishing techniques. The suggested technique's initial stage is deleting hyperlinks from the web page in question and bringing in relevant local web pages.

For phishing detection applications, Zabihimayvan and Doran used Fuzzy Rough Set (FRS) theory to pick significant characteristics from the dataset \cite{a32}. Rough Set (RS) theory is extended into the Fuzzy Rough Set (FRS) theory. Websites A and B are both phishing websites. Their features A and B have the same value, so RS is a method to find a decision boundary by calculating the equality of each data point based on certain features and the same classes, such as websites A and B being both phishing websites and their features A and B having the same value. RS is appropriate for the original dataset used in this paper, in which the features are used as a discrete value, i.e., they are a collection of 1, 0, -1 elements. However, after the dataset executes the nominalization process, the value of the feature is transferred to a continuous number from 0 to 1, and the FRS strategy is applied.

In 2021, El-Rashidy proposed a new method for selecting characteristics for an online phishing detection model \cite{ar23}. There are two parts to the feature selection process. The first phase determines the impact of each feature's absence by training the random forest model on a new dataset with one feature removed. A feature queue graded from high to low accuracy is created after removing each element in the loop. The model is then trained and tested by starting with one feature, adding a new feature from the ranking feature list each time to create a dataset, calculating the accuracy of each time, and ultimately finding the feature subset with the most excellent accuracy. This strategy is used to find the most useful feature subset. However, because each new dataset must go through the algorithm training and testing procedure, there is a lot of computational complexity and time involved.

Yang et al. \cite{ar18} proposed a new technique for phishing detection based on an inverted matrix online sequential over-learning machine that characterizes a website using three characteristics. To reduce matrix inversion, they employed the Sherman Morrison Woodbury equation. They introduced the online queue extreme learning machine to modify the training model.

Subasi and Kremic \cite{ar19} also developed a system for detecting phishing websites. They classified websites as legitimate or phishing using a variety of machine learning algorithms. They presented two types of learners (Adaptive Boosting (AdaBoost) and Multiboost) to help with anti-phishing and attack detection.

For phishing and Botnet attack detection, De La Torre Parra et al. \cite{ar20} suggested a cloud-based deep learning architecture. The model consists of two basic security mechanisms Distributed Convolutional Neural Network (DCNN) model that works together to detect phishing and application layer Distributed Denial of Service (DDoS) attacks and a cloud-based temporary long short-term memory (LSTM) network model hosted on the backend to detect botnet attacks and feed CNN to detect phishing attacks.

The primary focus of Yi et al. \cite{ar21} was developing a deep learning framework for detecting phishing websites. The researchers created two types of web phishing features original features and interactive features. These characteristics are employed in a Deep Belief Networks-based detection model (DBN). The DBN-based detection model produced promising results during experiments utilizing natural IP streams.

Wei et al. \cite{ar22} suggested a lightweight deep learning technique for detecting fraudulent URLs, allowing for a real-time and energy-efficient phishing detection system. They demonstrated that the suggested technology could identify phishing in real time using an energy-efficient integrated single board computer using website URLs. Table \ref{tab:my-table} depicts other existing machine learning models for phishing detection.

In this paper, we have developed and assessed web phishing detection models using Recurrent Neural Networks such as LSTM and GRU to achieve maximum accuracy and precision without compromising inference time for detecting malicious websites on small devices.

\begin{table*}
\centering
\caption{Existing Machine Learning Models for Phishing Detection}
\label{tab:my-table}
\resizebox{2\columnwidth}{!}{%
\begin{tabular}{cccccc} 
\toprule
Model or Algorithm & Type & Dataset & Challenges & Limitations & Accuracy \\ 
\midrule
Random Forest (RF\_1) \cite{tb1}& Single & ISCXURL-2016 & \begin{tabular}[c]{@{}c@{}}Achieved high\\ accuracy and low\\ response time\\ without relying on\\ third-party services\\ and using limited\\ features extracted\\ from a URL.\end{tabular} & \begin{tabular}[c]{@{}c@{}}Neither used multiple\\ different datasets to\\ train the model,\\nor compared the results\\ or evaluated the\\ robustness of\\ the model.\end{tabular} & 99.57\% \\
Adaboost \cite{tb2}& Single & \begin{tabular}[c]{@{}c@{}}Websites (PhishTank,\\ MillerSmiles, Google\\ Search), size of the\\ dataset not\\ mentioned, each\\ instance has\\ 30 features.\end{tabular} & \begin{tabular}[c]{@{}c@{}}The proposed model\\ used Weka 3.6,\\ Python, and\\ MATLAB.\end{tabular} & \begin{tabular}[c]{@{}c@{}}Didn't compare the results\\ and evaluate the\\ robustness of\\ the model.\end{tabular} & 98.30\% \\
Random Forest (RF\_2) \cite{tb3}& Single & \begin{tabular}[c]{@{}c@{}}Websites (PhishTank,\\OpenPhish, Alexa,\\online payment gateway)\\5223 instances, 2500\\phishing URLs,\\2723 legitimate URLs.\\Has 20 features.\end{tabular} & \begin{tabular}[c]{@{}c@{}}20 features are manually\\extracted, some features\\need to be obtained\\by calling a third-party\\service, and some\\features need to\\parse the website’s\\HTML source code.\end{tabular} & \begin{tabular}[c]{@{}c@{}}Did not use multiple\\different datasets to train\\the model, compare the\\results, or to evaluate\\the robustness of the\\model. The experimental\\dataset is small.\end{tabular} & 99.50\% \\
% \begin{tabular}[c]{@{}c@{}}LBET (logistic\\ regression + extra\\ tree) \cite{tb4}\end{tabular} & Hybrid & UCI & \begin{tabular}[c]{@{}c@{}}\\Combined meta-learning\\ algorithms and extra\\ trees to achieve high\\ accuracy and low\\ false-positive rate.\end{tabular} & \begin{tabular}[c]{@{}c@{}}Insufficient data\\ sources and lack of a\\ feature extraction\\ process.\end{tabular} & 97.57\% \\
PSL$^1$ + PART  \cite{tb5}& Hybrid & \begin{tabular}[c]{@{}c@{}}Websites (PhishTank,\\ Relbank)\\ 30,500 original\\ instances,\\ 20,500 phishing\\ URLs,\\ 10,000 legitimate\\ URLs. Has 18 features.\end{tabular} & \begin{tabular}[c]{@{}c@{}}\\Extracted 3000\\ comprehensive\\ features and applied\\ different set of\\ parameters to ML\\ models to compare\\ the experimental\\ results.\end{tabular} & \begin{tabular}[c]{@{}c@{}}The legitimate URLs\\ in the dataset are all\\ related to banks, and\\ some features are\\ limited to e-banking\\ websites.\end{tabular} & 99.30\% \\
\begin{tabular}[c]{@{}c@{}}Recurrent Neural\\ Network (RNN) +\\ Convolutional\\ Neural Network\\ (CNN) \cite{tb6}\end{tabular} & Deep learning & \begin{tabular}[c]{@{}c@{}}Websites (PhishTank,\\ Alexa, etc),\\ 490,408 instances,\\ 245,385 phishing URLs,\\ 245,023 legitimate URLs.\end{tabular} & \begin{tabular}[c]{@{}c@{}}\\Dataset is large-scale.\\ The first one to use\\ deep learning\\ model to detect\\ malicious URLs.\end{tabular} & \begin{tabular}[c]{@{}c@{}}The maximum length\\ of the URL is\\ 255 characters.\\ The phishing website\\ URLs did not have\\ relevant semantics.\end{tabular} & 95.79\% \\
\begin{tabular}[c]{@{}c@{}}Random forest +\\ Neural network +\\bagging \cite{tb7}.\end{tabular} & Hybrid & UCI & \begin{tabular}[c]{@{}c@{}}\\No previous research\\ focuses on using a\\ feedforward NN \\ensemble learning.\end{tabular} & \begin{tabular}[c]{@{}c@{}}Insufficient data\\ sources and lack of a\\ feature extraction\\ process.\end{tabular} & 97.40\% \\
\begin{tabular}[c]{@{}c@{}}Auto encoder +\\ NIOSELM \cite{tb8}\end{tabular} & Hybrid & \begin{tabular}[c]{@{}c@{}}Websites (PhishTank,\\ Alexa, DMOZ),\\ 60,000 legitimate URLs,\\ 5000 phishing URLs.\\Has 56 features.\end{tabular} & \begin{tabular}[c]{@{}c@{}}The dataset is\\ imbalanced.\end{tabular} & \begin{tabular}[c]{@{}c@{}}The detection\\ accuracy may not be\\ the best compared\\ with the existing\\ methods.\end{tabular} & 94.60\% \\
\bottomrule
\end{tabular}%
}
\end{table*}

\section{Methodology}
\label{Sec:Methodology}

\textbf{Long Short-Term Memory} \cite{article26,a36} was introduced to address the long-term dependency problem. LSTM has a hidden state in which an LSTM block would output and an internal cell that maintains the information across a temporal context. The cell stores the long-term information, and the LSTM can erase, write and read information from that cell based on whatever a context defines. In step t, there is a hidden state $h_t$ and a cell state $C_t$.
The selection of the information erased or written, or read is controlled by three corresponding gates each for erasing, writing, and reading respectively.
These gate values are dynamic and are learned and computed based on the input at a particular time step and the hidden state from the previous time step.

\begin{figure}
\centering
\includegraphics[width=0.5\textwidth]{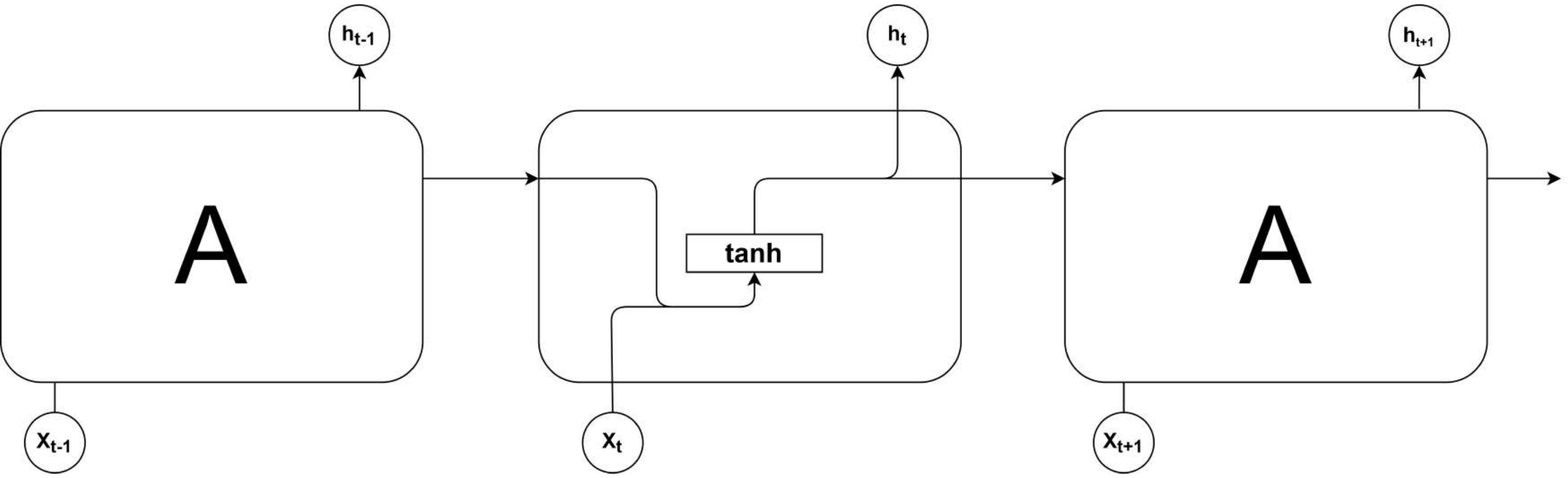}
\caption{Vanilla RNN architecture.}
\label{FIG:Vanilla_RNN_architecture.}
\end{figure}

Figure. \ref{FIG:Vanilla_RNN_architecture.} is a single neural network with inputs $X_{t-1}$, $X_{t}$, $X_{t+1}$ at $t-1$, $t$ and $t+1$ that has a $tanh$ activation function. The input at a particular time step $h_t$, as well as $h_{t-1}$ (output of the previous RNN block), which is provided as input, and $tanh$ is applied that gives us $h_t$. It is given as output and input for the next RNN block at the next time step. LSTM has a similar structure, and it is a chain of repeating modules where the same block applies at every step. The structure of each block contains four different layers that interact with each other and are not recurrent layers.

\begin{table*}[h]
\centering
\caption{Variables of LSTM Unit}
\resizebox{0.85\linewidth}{!}{%
\begin{tabular}{ll} 
\toprule
\multicolumn{2}{c}{Variables of LSTM Unit}  
\\ \midrule 
$x_t \in \mathbb{R}^{d}$     & input vector to LSTM Unit                  \\
$f_t \in (0,1)^{h}$          & forget gate's activation vector                  \\
$i_t \in (0,1)^{h}$          & input/update forget gate's activation vector    \\
$o_t \in (0,1)^{h}$          & output forget gate's activation vector             \\
$h_t \in (-1,1)^{h}$         & hidden state vector also known as output vector              \\
$\tilde{c}_{t} \in (-1,1)^h$ & cell input activation vector              \\
$c_t \mathbb{R}^{h} $        & cell state vector              \\
$W \in \mathbb{R}^{h \times d }, U \in \mathbb{R}^{h \times d }~and~b \in \mathbb{R}^h$ & weight matrices and bias vector parameters \\
\bottomrule
\end{tabular}
}
\end{table*}

\begin{figure}
\centering
\includegraphics[width=0.5\textwidth]{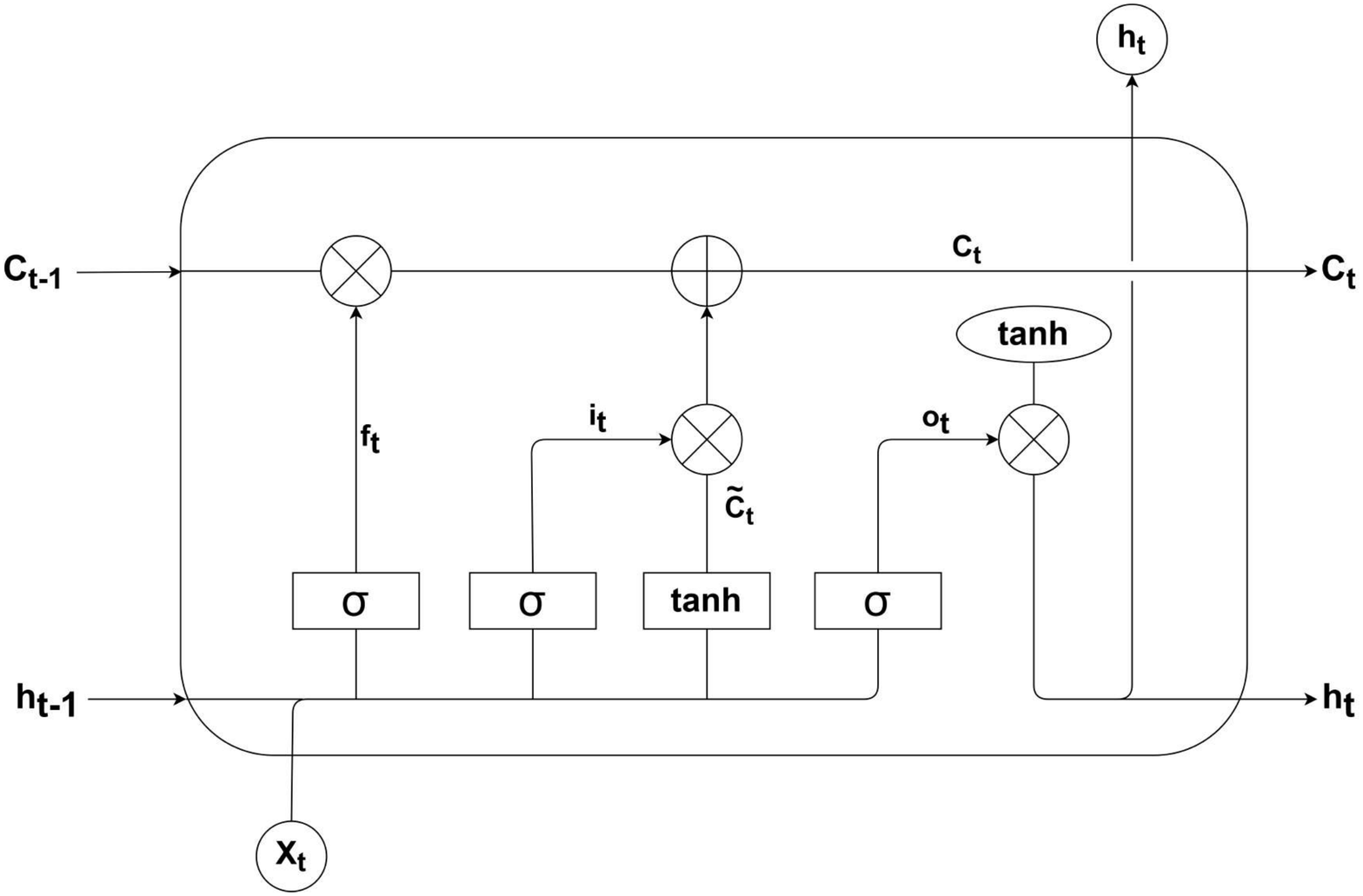}
\caption{LSTM architecture.}
\label{FIG:LSTM_architecture}
\end{figure}

The sigmoid function is used in these three blocks because the output should lie between 0 \& 1. The cell state ($C_t$) can remove and add information to the cell state, which is regulated by gates. The following points show the working of LSTM.

\begin{itemize}
    \item Forget gate layer uses a sigmoid function to decide the information to be thrown away from the cell state.
    	\begin{equation} f_t=\sigma(U_f h_{t-1}+ x_t W_f+b_f) \end{equation} 
    
    \item Input gate layer(sigmoid function) decides the values to be updated, and the $tanh$ function creates the new vector value, which is added to the state.
    
        \begin{equation} i_t=\sigma(x_t W_i + h_{t-1} U_i + b_i) \end{equation}
        \begin{equation} \tilde{C}_t=tanh(x_t W_C + h_{t-1} U_c + b_c) \end{equation}
    
    \item Multiply the old cell with the cell which contains information that needs to be dropped, and sum it with the dot product of the result from the previous step.
    
    \begin{equation} C_t=\sigma(f_t \times C_{t-1}+i_t \times \tilde{C}_t) \end{equation}
    
    \item A filtered version of output is formed in the cell state; run the sigmoid layer to decide the part of the cell states for output. Apply $tanh$ activator to the cell state and multiply it with the output of the sigmoid gate, where the value lies between 0-1.
    
    \begin{equation} o_t=\sigma(x_t W_o + h_{t-1} U_o + b_o) \end{equation}

    \begin{equation} h_t=o_t\times tanh(C_t) \end{equation}
\end{itemize}

\begin{figure}
\centering
\includegraphics[width=0.5\textwidth]{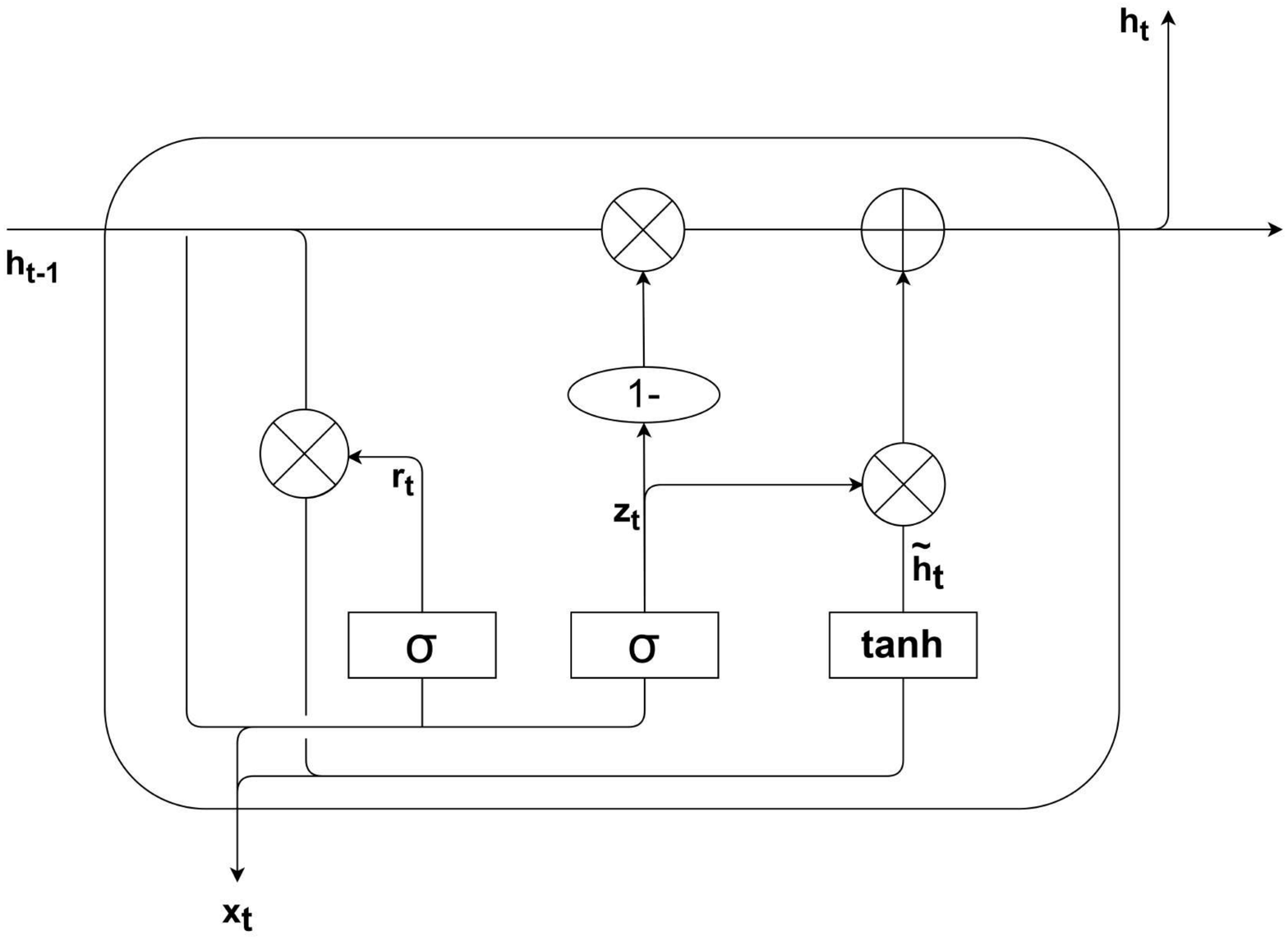}
\caption{GRU architecture.}
\label{FIG:GRU_architecture}
\end{figure}

\textbf{Gated Recurrent Unit} \cite{article27} was proposed as an alternative to LSTM, which is more straightforward; it combines forget and input gates into a single update gate as well as merges the cell state and hidden state. Figure \ref{FIG:GRU_architecture} shows GRU architecture.

The update gate controls what parts of the hidden state are updated and preserved.

\begin{equation} z_t=\sigma(x_t W^z+h_{t-1} U^z+ b_z) \end{equation}

Reset gate controls which parts of the previous hidden state are used to compute the fresh subject.
% \begin{equation} r_t=\sigma(x_t W^r+h_{t-1} U^r+ b_r) \end{equation}
New hidden state(reset gate) content selects useful parts of the previous hidden state. It uses the current input to compute new hidden content. 
\begin{equation} \tilde{h}_t=tan(r_t \times h_{t-1}U + x_t W + b) \end{equation}

The hidden state (update gate) controls what is kept from the previous hidden state and what is updated to new hidden state content.

\begin{equation} h_t=(1-z_t)\times \tilde{h}_{t}+z_t\times h_t-1 \end{equation}

If reset gate is set to all 1’s and update gate to all 0’s then in reset gate $h_t$ will be $tanh(W * [h_{t-1}], X_t)$ and in update gate, $h_t$ becomes, $h_{t-1}$ respectively. Input and forget gates of LSTM are coupled by an update gate in GRU; the reset gate in GRU is applied directly to the previous hidden state.

\begin{table}[h]
\centering
\caption{Variables of GRU Unit}
\resizebox{0.85\linewidth}{!}{%
\begin{tabular}{ll} 
\toprule
\multicolumn{2}{l}{Variables of GRU Unit}  
\\ \midrule 
$x_t$                                       & input vector                   \\
$h_t$                                       & output vector                  \\
$\tilde{h}_{t}$          & candidate activation vector    \\
$z_t$                                       & update gate vector             \\
$r_t$                                       & reset gate vector              \\
$W, U~and~b$ & parameter matrices and vector  \\
\bottomrule
\end{tabular}
}
\end{table}

\section{Proposed Model}

\subsection{Preparation of Dataset}
\label{dataprep}
For effective detection of malicious URLs, the dataset should contain recent URLs which are malicious, for recognizing fresh features to train the model. Attackers will change the production of phishing links by anti-phishing regulations and procedures that have been released. Anti-phishing models and algorithms must also be improved based on new phishing data. Furthermore, the training dataset's quantity and validity significantly impact the performance of machine learning-based solutions. The performance of deep learning models increases with the variety of content in the training dataset. Hence, it is advised that phishing URLs and legitimate URLs should be extracted from data repositories. The dataset used in this paper consists of numerous legitimate and malicious URLs which are taken from PhishTank, OpenPhish, and Common Crawl. It consists of 46839 instances, and we merely looked into the text in the URL and extracted features to train the model. The dataset was split into 75\% and 25\% for training and testing, respectively.

\subsection{Network Architecture and Training Parameters}
\label{training}

\begin{figure}[h]
\centering
\includegraphics[width=0.5\textwidth]{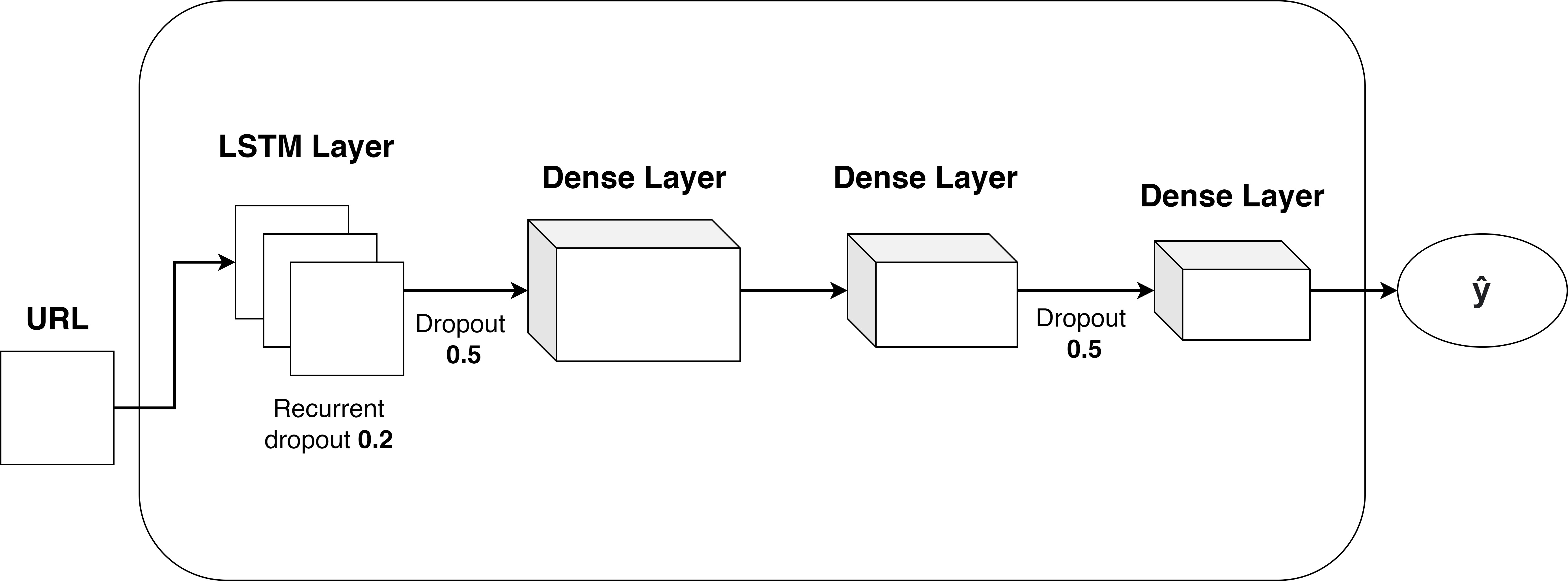}
\caption{Comparison of Accuracy at different epochs.}
\label{FIG:LSTM_ARC}
\end{figure}

\begin{figure}[h]
\centering
\includegraphics[width=0.5\textwidth]{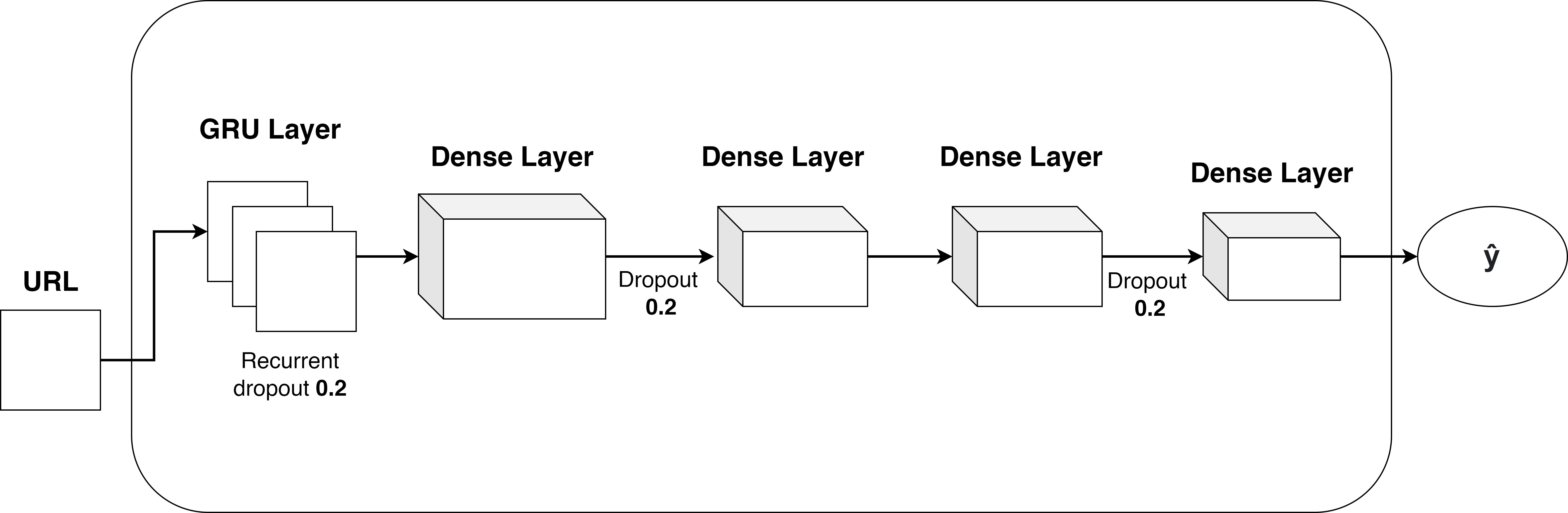}
\caption{Comparison of Accuracy at different epochs.}
\label{FIG:GRU_ARC}
\end{figure}
All the features captured from the URL are fed to the LSTM layer with an orthogonal recurrent initializer. The embedded matrix is used as weights, further combined with dense layers with a sigmoid activation function. A dropout of 0.5 is added between the LSTM layer and dense layer. Similarly, the features are captured from the URL and fed to the GRU layer with an orthogonal recurrent initializer and sigmoid recurrent activation, combined with dense layers with a softmax activation function. A dropout of 0.2 is added between the dense layers. Figure \ref{FIG:LSTM_ARC} and \ref{FIG:GRU_ARC} depict the network architecture of Phish-Defence of LSTM and GRU, respectively. The trained model is converted to a tflite model for producing faster inference time on small edge devices like Raspberry Pi.

The model is trained for 40 epochs with a batch size of 500 using the $tanh$ activation function for the LSTM model and the sigmoid activation function for the GRU model. Binary cross entropy loss with Adam optimizer is attached to the dense layer for classifying legitimate and malicious URLs in both LSTM and GRU models. Binary cross-entropy loss is defined as

\begin{equation}loss = - \frac{1}{N}\sum_{i = 1}^{N}y_{i} \bullet log\left( p\left( y_{i} \right) \right) + \left( 1 - y_{i} \right) \bullet log\left( 1 - p\left( y_{i} \right) \right)\end{equation}

\section{Results}
\label{Sec:results}
As described in Section \ref{training} the network architectures were trained with a batch size of 500 with 70 training steps for 40 epochs, and the training and testing loss started oscillating. The Adam optimization method was utilized with an initial learning rate of 1e-3. This learning rate was decreased by a learning rate scheduler, which checks for lack of reduction in training loss for five epochs and decreases the learning rate by a factor of 0.1. The reduction was stopped when the learning rate was reduced to an absolute minimum of 1e-5. Early stopping was set at 6 epochs, which ends the training process when the training loss oscillates for more than 6 epochs, yielding a maximum validation accuracy of 90.50\% for the LSTM model.

\begin{table}[h]
\centering
\caption{Results of Phish-Defense}
\label{TBL:RES}
\resizebox{\columnwidth}{!}{%
\begin{tabular}{@{}lllll@{}}
\toprule
Model                                                                                                               & Precision & Inference time & F-Score & Recall \\ \midrule 
Random Forest (RF\_1) \cite{tb1}                                                                                                       & 98.30      & 2.30               & 99.01   & 99.65  \\ \midrule
Adaboost \cite{tb2}                                                                                                            & 99.00        & 2.80               & 98.80    & 98.60   \\ \midrule
Random Forest (RF\_2) \cite{tb3}                                                                                                       & 99.40      & 1.95                & 99.40    & -      \\ \midrule
PSL ${^1}$ + PART \cite{tb5}                                                                                                          & 98.70      & 3.10               & 99.00      & 99.30   \\ \midrule
\begin{tabular}[c]{@{}l@{}}Recurrent Neural\\ Network (RNN) +\\ Convolutional\\ Neural Network\\ (CNN) \cite{tb6}\end{tabular} & 97.33     & 2.03               & 95.52   & 93.78  \\ \midrule
\begin{tabular}[c]{@{}l@{}}Random forest +\\ Neural network +\\ bagging \cite{tb7}\end{tabular}                                & 96.00        & 1.90               & 97.00      & 98.10   \\ \midrule
\begin{tabular}[c]{@{}l@{}}Auto encoder +\\ NIOSELM \cite{tb8}\end{tabular}                                                    & 97.80      & 2.20               & 95.52   & 98.30   \\ \midrule
PD-LSTM                                                                                                       & 91.02      & 0.60                & 96.40    & 97.54      \\ \midrule
PD-GRU                                                                                                       & 99.08      & 0.53                & 99.24    & 98.97      \\ \midrule
\end{tabular}%
}
\end{table}

\begin{figure}[h]
\centering
\includegraphics[width=0.5\textwidth]{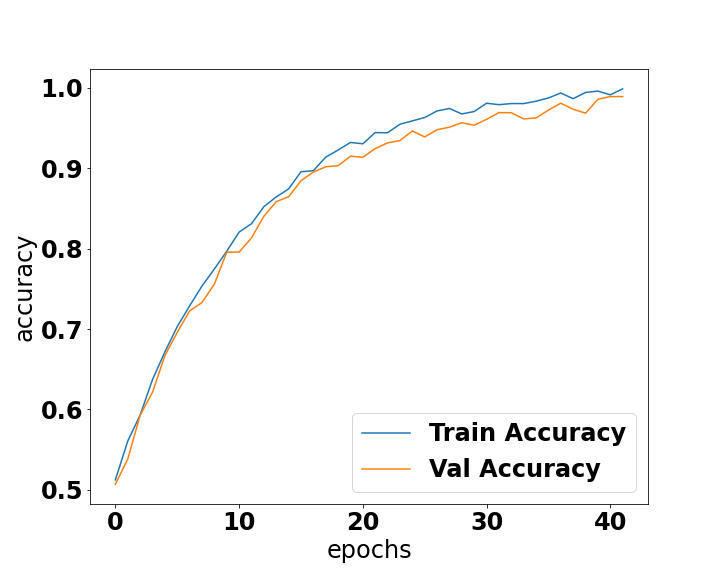}
\caption{Accuracy vs Epochs for GRU model.}
\label{FIG:accuracy_epochs}
\end{figure}

\begin{figure}[h]
\centering
\includegraphics[width=0.5\textwidth]{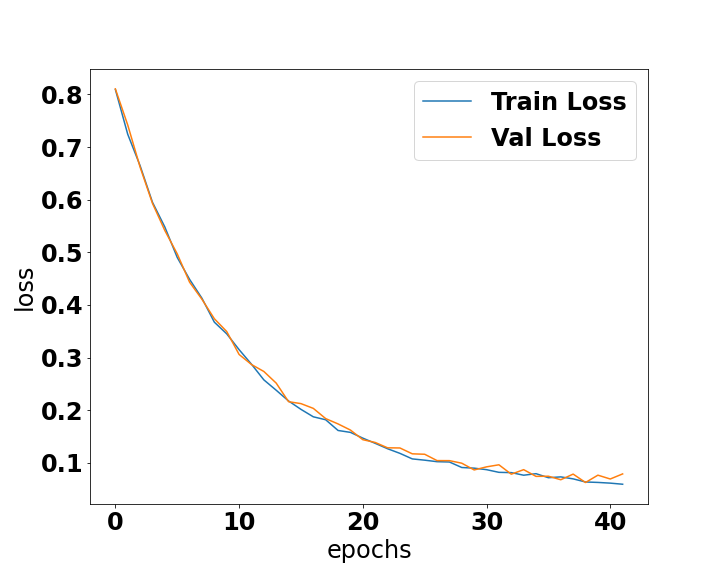}
\caption{Loss vs Epochs for GRU model.}
\label{FIG:loss_epochs}
\end{figure}

\begin{figure*}[h]
\centering
\includegraphics[width=\textwidth]{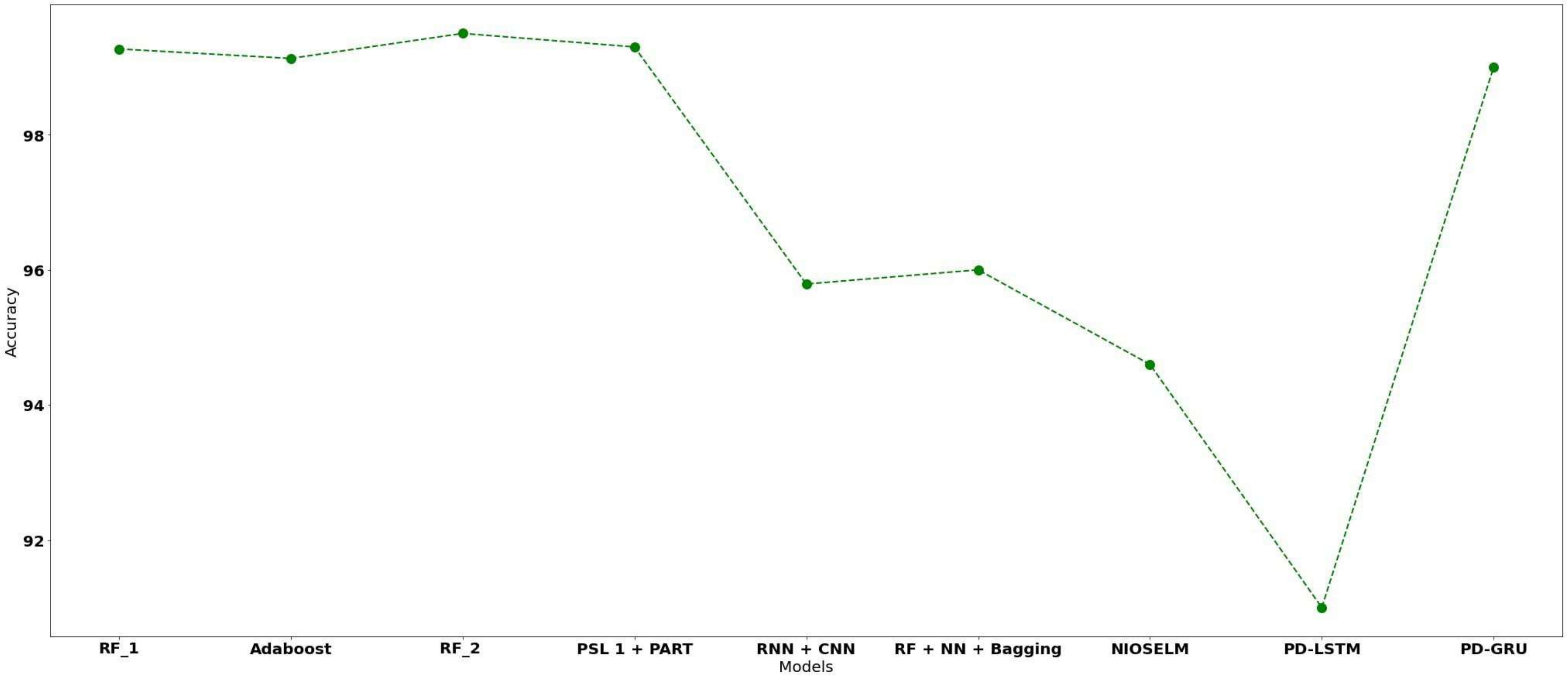}
\caption{Comparison of accuracy with existing models.}
\label{FIG:accuracy}
\end{figure*}

\begin{figure*}[h]
\centering
\includegraphics[width=\textwidth]{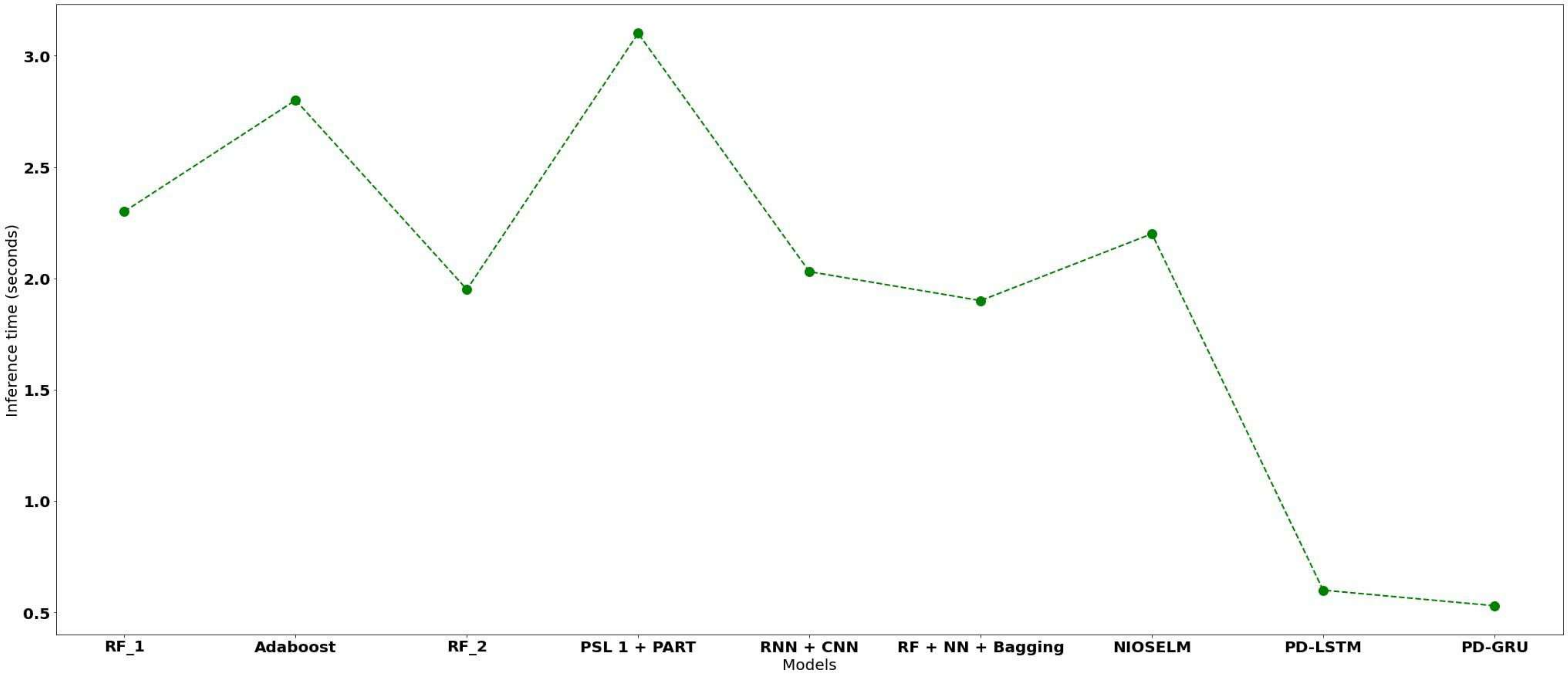}
\caption{Comparison of inference time with existing models.}
\label{FIG:inference}
\end{figure*}

Later, we built and trained a fresh RNN model using GRU and achieved a state-of-the-art accuracy of 98.51\%. Figure \ref{FIG:accuracy_epochs} and \ref{FIG:loss_epochs} shows the accuracy vs epochs and loss vs epochs graph for Gated Recurrent Unit model. This accuracy was achieved using the softmax activation function with a 0.25 dropout. After performing a significance test, the resulting trained model was found to perform with an inference time of 0.8 seconds on Raspberry Pi-4. Figure \ref{FIG:accuracy} and \ref{FIG:inference} illustrate the accuracy and inference time in the form of a line graph.  
Table \ref{TBL:RES} shows the results of Phish-Defense compared to other machine learning algorithms.

\section{Conclusion}
\label{Sec:conclusion}

This study recommended a web phishing detection approach using deep recurrent neural networks to predict phishing websites. In phishing site prediction, the most influential features and optimal weights of website features were tokenized, and these features were used to train the RNN to make better predictions of phishing websites. The experimental results demonstrated that the proposed phishing websites prediction approaches based on RNN increased the classification performance using fewer features. Additionally, the proposed method approaches produced the competitive classification accuracy of phishing websites among all other classifiers with all the feature selection methods used in this study with the lowest inference time. Thus, the proposed approach based on RNNs can also be used on smaller devices like Raspberry-Pi and Arduino to successfully predict phishing websites to contribute and provide more confidence for online commerce and business customers.

\section*{Acknowledgments}
\label{Sec:acknowledgments}

This research is carried in the Artificial Intelligence and Robotics (AIR) Research Centre, VIT-AP University. We also thank the management for motivating and supporting AIR Research Centre, VIT-AP University in building this project. \\

\bibliographystyle{ieeetr}
\bibliography{output}
\end{document}